# Very high thermoelectric power factor in a $Fe_3O_4/SiO_2/p$-type Si(100) heterostructure

Z. Viskadourakis,[1,2,†] M. L. Paramês,[3] O. Conde,[3] M. Zervos,[1,4] and J. Giapintzakis[1,4,a]

[1]*Department of Mechanical and Manufacturing Engineering, School of Engineering, University of Cyprus, 75 Kallipoleos Avenue, P.O. Box 20537, 1678 Nicosia, Cyprus*

[2]*Department of Materials Science and Technology, University of Crete, P.O. Box 2208, 71003 Heraklion, Crete, Greece*

[3]*Department of Physics, University of Lisbon and ICEMS, Campo Grande Ed. C8, 1749-016 Lisbon, Portugal*

[4]*Nanotechnology Research Unit, School of Engineering, University of Cyprus, 75 Kallipoleos Avenue, P.O. Box 20537, 1678 Nicosia, Cyprus*

[†]*Current address: Crete Center for Quzntum Complexity and Nanotecnology, P. O. Box 2208, GR7-1003 Heraklion, Greece*

[a] *Corresponding author's e-mail: giapintz@ucy.ac.cy*

**Abstract**

The thermoelectric and transport properties of a $Fe_3O_4/SiO_2/p$-Si(100) heterostructure have been investigated between 100 and 300 K. Both Hall and Seebeck coefficients change sign from negative to positive with increasing temperature while the resistivity drops sharply due to tunneling of carriers into the *p*-Si(100). The low resistivity and large Seebeck coefficient of Si give a very high thermoelectric power factor of 25.5mW/$K^2$m at 260K which is an underestimated, lower limit value and is related to the density of states and difference in the work functions of $Fe_3O_4$ and Si(100) that create an accumulation of majority holes at the p-Si/$SiO_2$ interface.

**Introduction**

Thermoelectricity has been investigated for about two hundred years since it can be used to convert heat to electricity through the Seebeck effect or for Peltier cooling. The efficiency of conversion is described by the dimensionless, thermoelectric figure-of-merit, $ZT = S^2T/\rho\kappa$, where $S$ is the Seebeck coefficient or thermopower, $\rho$ is the electrical resistivity, $T$ is the absolute temperature, and $\kappa$ is the total thermal conductivity which consists of the electronic and lattice thermal conductivities. A high thermoelectric figure-of-merit may be achieved by reducing $\kappa$ and/or by enhancing the thermoelectric power factor, $P = S^2/\rho$. To this end, extensive investigations have been carried out over the past six decades, but progress has been limited due to the interdependenceof $S$, $\rho$, and $\kappa$ [1–7]. This constraint may be surpassed by the use of low dimensional semiconductors [2–8] such as semiconductor nanowires. Recently, it was shown that Si nanowires with diameters of 20–300 nm had $S$ and $\rho$ that were close to the values of bulk Si but those with diameters of 50 nm exhibited a one-hundred fold reduction in thermal conductivity giving very high $ZT$ [3]. On the other hand, large power factors $P$ can be obtained by tuning the electronic structure in order to obtain a narrow density of states (DOS) around the Fermi level as in the case of narrow energy gap semiconductors with correlated bands [9]. A huge thermoelectric power factor of $P = 200$ mW/K$^2$m was measured by Sun *et al.*[10] in FeSb$_2$ which is about ten times larger than the previous highest $P = 23.5$ mW/K$^2$m in YbAgCu$_4$ and outweighs significantly the thermoelectric power factor $P = 4$ mW/K$^2$m of optimized Bi$_2$Te$_3$ [11]. The large value of $P$ was shown to be electronic in origin and structuring on the nm scale was suggested as a way to reduce $\kappa$ and obtain very high $ZT$'s [12, 13].

Here we describe an Fe$_3$O$_4$ / SiO$_2$ / $p$-type Si(100) heterostructure [14, 15] which exhibits thermoelectric power factors as high as 25.5 mW/K$^2$m at 260 K. These state-of-the-art values of $P$ are a direct consequence of the density of states of Fe$_3$O$_4$ and differences in the work functions of Fe$_3$O$_4$ and $p$-type Si(100) which lead to band bending and an accumulation layer of holes at the Si / SiO$_2$ interface, rather than an inversion layer of electrons, which is responsible for a low $\rho$ and high $S$. We note that these values of $P$ are underestimated, lower limit, values and that the use of Si nanowires combined with Fe$_3$O$_4$ is interesting and promising for the realization of high thermoelectric figure of merit devices.

**Experimental Details**

Polycrystalline Fe$_3$O$_4$ thin films with randomly oriented grains were grown on $p$-type Si(100) substrates at 210 $^o$C using pulsed laser deposition (PLD). Details of the growth conditions and parameters are described in detail elsewhere [14]. The electric transport and thermoelectric properties described below were obtained from a 200 nm $Fe_{2.99}O_4$ film whose stoichiometry was determined via conversion electron Mössbauer spectroscopy (CEMS) at room temperature. The *p*-type Si(100) had a native $SiO_2$ layer with an estimated thickness of ~ 2 nm while CEMS indicated that there was no interdiffusion of Fe ions at the $Fe_3O_4$ / $SiO_2$ interface. The sheet resistance ($R_S$) and Hall voltage ($V_H$) were measured between 100 and 300 K, using the van der Pauw method. The Seebeck coefficient was measured as a function of temperature using the steady state technique in the in-plane configuration shown in Fig. 1.

**Results and Discussion**

We will begin by considering the electrical transport properties of the $Fe_3O_4$/ $SiO_2$ / *p*-type Si(100) heterostructure and then consider its thermoelectric properties. The temperature dependence of the sheet resistance $R_S$ is shown as an inset in Fig. 2. The most notable feature is the considerable drop in the resistance occurring at ≈ 230 K. Similar R(T) dependence has recently been reported in $Fe_3O_4$ / $SiO_2$ / Si [16, 17], $Fe_3C$ / $SiO_2$ / Si [18], and other M / $SiO_2$ / Si structures, where M stands for very-thin metallic films [19] and has been attributed to a switching of conduction from the metal to the Si inversion layer of holes at the Si / $SiO_2$ interface via thermally assisted tunneling. More specifically, these metals have been deposited on *n*-type Si. As we will show later, we do not have an inversion layer of holes but rather an accumulation layer of holes at the Si/$SiO_2$ interface since we are dealing with *p*-type Si(100).

The Hall voltage ($V_H$) of the $Fe_3O_4$ / $SiO_2$ / *p*-type Si(100) heterostructure is shown in Fig. 3 as a function of temperature and exhibits a change of sign from negative to positive at 213K which is evidence that two types of charge carriers coexist in the system. The charge carriers in the $Fe_3O_4$ [20] are electrons which give a negative $V_H$ whereas the holes of the p-type Si(100) are responsible for a positive $V_H$. It is worthwhile pointing out that below the Verwey transition temperature ($T_V$) of 120 K, there is an abrupt, negative, increase in $V_H$, which is shown as an inset in Fig. 2, corresponding to a drastic reduction in the carrier density of electrons and the metal-to-insulator transition of $Fe_3O_4$. Similar to the temperature dependence of $V_H$ we find that the in - plane magnetoresistance defined as [R(H)-R(0)]/R(0)) and measured with H ⊥ Si(100) up to 6 T is negative at low temperatures and comparable to other values reported for polycrystalline Fe3O4 films grown on MgO [21] and Si [22, 23] while it becomes positive above 200 K. The negative TMR is attributed to a current flow of electrons in the Fe$_3$O$_4$ film while the positive TMR at higher temperatures is attributed to hole current flow in the *p*-type Si(100). This behavior of transverse magnetoresistance (TMR) is consistent with that reported for Fe$_3$O$_4$ / SiO$_2$/ *n*-type Si heterostructures [24] as well as for other M-SiO$_2$-Si structures [19, 24].

Before considering the thermoelectric properties of the 200 nm Fe$_3$O$_4$ / 2 nm SiO$_2$/ *p*-type Si(100) heterostructure it is worthwhile considering the energy band diagram of this metal-oxide-semiconductor (MOS) structure which is shown in Fig. 1 by taking into account the proper boundary conditions. The resistivity of the *p*-type Si(100) substrate was 8.2 Ωcm and had a hole concentration of p = 1.8x10$^{15}$cm$^{-3}$ determined from Hall effect measurements at room temperature. We find from p = N$_V$exp{E$_V$ - E$_F$ / kT} that E$_F$ - E$_V$ = 0.25 eV, where N$_V$(Si) = 2x10$^{19}$cm$^{-3}$ at 300K. The work function of *p*-type Si(100) is dependent on the doping level and is given by $\varphi_S = \chi$ + (E$_C$ - E$_F$), i.e., $\varphi_S$ = 4.92 eV since the electron affinity $\varphi_{Si}$ = 4.05 eV and E$_C$ - E$_F$ = 0.87 eV. On the other hand, the work function of clean Fe$_3$O$_4$ is $\varphi_M$ = (5.5 ±0.05) eV [25] and since the energy gap of SiO$_2$ is ≈9.0 eV with $\chi_{SiO2}$ = 1.0 eV we obtain a barrier height $\varphi_B$ = 4.5 eV at the Fe$_3$O$_4$ / SiO$_2$ interface. From the above we find that the work function difference between the metal and semiconductor ($\varphi_M$ - $\varphi_S$ = $\varphi_{MS}$ = 0.58 eV) is positive, which will produce an upwards band bending at the Si/SiO$_2$ interface leading to an accumulation of holes in a quasi triangular potential well. Here we ought to mention that the lowest limit for the work function of $\varphi_M$ corresponding to Fe$_3$O$_4$ which is not clean is $\varphi_M$ = 4.3 eV [26] in which case $\varphi_M$ - $\varphi_S$ = $\varphi_{MS}$ = - 0.62 eV which will result into a downward band bending at the Si/SiO$_2$ interface. However, this is not sufficient enough to cause an inversion layer of electrons at the Si/SiO$_2$ interface since the Fermi level resides almost 0.9 eV below the conduction band edge deep in the p-type bulk. The overall band-bending and depth of the quasi triangular potential well are essentially dictated by the difference in the work functions $\varphi_M$ (Fe$_3$O$_4$) and $\varphi_S$(Si) but also any charges that are present at the Fe$_3$O$_4$ / SiO$_2$ or SiO$_2$ / Si interfaces. Since we are dealing with nearly stoichiometric Fe$_{2.99}$O$_4$ and since there was no interdiffusion of Fe ions at the Fe$_3$O$_4$/ SiO$_2$ interface, we do not expect that the Fe ions located at the Fe$_3$O$_4$ / SiO$_2$ or SiO$_2$ / Si interfaces will govern the electronic properties of this heterostructure. We take the work function of Fe$_3$O$_4$ to be $\varphi_M$ = 5.5 eV similar to the value quoted by Qu *et al.* [27] who investigated the vertical or cross-plane electrical transport properties of a Fe$_3$O$_4$ / SiO$_2$ / *n*-Si(100) heterojunction grown by PLD. However, in the case of the Fe$_3$O$_4$ / SiO$_2$ / *p*-Si(100) heterostructure considered here we have $\varphi_M$ - $\varphi_S$ > 0 and $\varphi_{MS}$ = 0.58 eV which is smaller than that of $\varphi_{MS}$ =1.0 eV in the case of Qu *et al.* [27] by virtue of the fact that we have used *p*-type Si(100). Despite the fact that $\varphi_{MS}$ is smaller it is still positive and this difference in work functions will produce an upward band bending at the Si / SiO$_2$ interface that will make the valence band edge cross the Fermi level since the latter remains close to the valence band edge far away from the interface. Therefore, the accumulation of holes is expected to be more pronounced in the case of Fe$_3$O$_4$ on $p$-type Si(100) as opposed to Fe$_3$O$_4$ on $n$-type Si(100) and holes remain the majority carrier in $p$-type Si which is extremely critical for a high S. In other words an accumulation of holes at the Si / SiO$_2$ interface of a Fe$_3$O$_4$ / SiO$_2$ / $n$-type Si(100) heterostructure is expected to lead to a smaller S.

In addition to the $p$-type Si(100) it is also necessary to consider the electronic structure of Fe$_3$O$_4$ after Camphausen *et al.* [28] who shows that t$_{2g}$ electrons occupy the minority spin down band which is located at the Fermi level giving rise to the half-metallic-like or semiconducting behavior of Fe$_3$O$_4$. We do not expect significant band bending to be supported by the Fe$_3$O$_4$ within the immediate vicinity of the Fe$_3$O$_4$ / SiO$_2$ hetero-interface and certainly no depletion by virtue of the continuity of the electric field which gives rise to a downward band bending in the conduction band of Fe$_3$O$_4$ as shown in Fig. 1. Moreover, we expect that the density of holes in the $p$-type Si(100) far away from the Si / SiO$_2$ interface will drop with decreasing temperature and the Fermi level will start moving towards the middle of the gap in the $p$-type Si bulk reducing $\varphi_S$ and increasing $\varphi_{MS}$. This is expected to increase the band bending at the Si / SiO$_2$ interface. In addition, according to the Verwey model, charge ordering below T$_V$ should open a gap at the Fermi energy resulting into the metal - insulator transition [29] so it is unlikely that the Fermi level will move upwards thereby reducing $\varphi_M$ of Fe$_3$O$_4$.

The temperature dependence of S for the Fe$_3$O$_4$ / SiO$_2$ / $p$-type Si(100) heterostructure is shown in Fig. 2(a). Seebeck coefficient data for a Fe$_3$O$_4$ / 1.5 nm AlO$_x$ / GaAs (100) heterostructure as well as for a magnetite single crystal [30] are also included for comparison. It is seen that the thermopower of the magnetite single crystal is small and negative in the entire measured temperature range (~ -60 $\mu$V/K) and the data for the Fe$_3$O$_4$ / 1.5 nm AlO$_x$ / GaAs (100) heterostructure follows the single crystal Fe$_3$O$_4$ closely. On the other hand, the S curve of the Fe$_3$O$_4$ / SiO$_2$ / $p$-Si(100) structure exhibits entirely different behavior. At temperatures below 120K the measured thermopower is negative as expected and then changes to being positive and remains constant up to 200 K. S further increases with increasing temperature and exhibits a large increase in the temperature range (200–300K). At higher temperatures S reaches quite large values (+ 900 $\mu$V/K), which are comparable to the S value of the $p$-type Si substrate ( +1230 $\mu$V/K at room temperature).

The Seebeck coefficient gives a clear indication of the majority carrier in a homogeneous semiconductor, and it is inversely proportional to the carrier density. At first sight one would be inclined to suggest that the value of $S$ will depend on the density of holes at the Si / $SiO_2$ interface and in the $p$-type Si(100) bulk which will be reduced by the negative contribution of the electrons residing in the $Fe_3O_4$. At high temperatures $S$ is dominated by the $p$-type bulk Si(100) and the accumulation layer of holes at the Si / $SiO_2$ interface. As the temperature is reduced we find that $S$ drops and then remains constant down to $T_V$ = 120 K. Considering that $S$ is constant and negative all the way between 120 and 300K for single crystal $Fe_3O_4$ one could simply argue that the reduction of $S$ comes about from an increase in the density of holes which will reduce the positive part of the Seebeck coefficient. However, here we are dealing with a $Fe_3O_4$ / $SiO_2$/ $p$-type Si(100) heterostructure where a temperature gradient is applied on top of the $Fe_3O_4$. Therefore, a potential difference will develop in the underlying $p$-type Si via tunneling of "hot" electrons from the $Fe_3O_4$ into the $p$-type Si(100) which is clearly suppressed at low temperatures. This tunneling is in fact responsible for the drastic reduction in the resistance above 250 K, i.e., electrons tunnel across the $SiO_2$ barrier and conduct along the Si / $SiO_2$ interface in a quasi triangular potential well whose spatial extent normal to the interface is of the order of a few tens of nm's. At low temperatures this tunneling is suppressed and so the contribution of the underlying $p$-type Si(100) towards $S$ is reduced.

The thermoelectric power factor $P$ of the $Fe_3O_4$ / $SiO_2$ / $p$-Si(100) heterostructure is shown as a function of temperature in Fig. 2(b). It is noted that the values of $\rho$ used in calculating $P$ were estimated using the measured $R_S$ and the thickness of the $Fe_3O_4$ for the entire temperature regime. Since for T > 210K carrier conduction is believed to be confined to the accumulation layer, which is expected to have a smaller thickness than the $Fe_3O_4$ film, the values of $\rho$ are overestimated in the temperature region of interest and thus the $P$ values correspond to lower-bound values. The maximum $P$ value (25.5 mW/K$^2$m) is obtained at ≈ 260K coinciding with the resistance drop and the change in Hall coefficient, whereas the room-temperature $P$ value is slightly lower (21.5 mW/K$^2$m). These $P$ values are among the highest ones reported to date and are caused by the low values of $\rho$ associated with tunneling of electrons from the $Fe_3O_4$ into Si but also the large values of $S$ from the underlying $p$-type Si(100) which are not expected to occur in a $Fe_3O_4$ / $SiO_2$ / $n$-type Si(100) heterostructure with an inversion layer of holes.

From the above it appears that the $Fe_3O_4$ provides an effective way of maintaining a high Seebeck coefficient of $p$-type Si(100) and low $\rho$ path thereby making the power factor large which would not be possible with any choice of metal. For example, Ni which has an intrinsic room temperature Seebeck coefficient of - 16 $\mu$V/K and a smaller work function of 5.1 eV gave a smaller but positive Seebeck coefficient of + 24 $\mu$V/K at room temperature as opposed to + 900 $\mu$V/K obtained with $Fe_3O_4$. The properties of the $Fe_3O_4$ / $SiO_2$ / *p*-type Si(100) heterojunction are interesting since Hicks and Dresselhaus [8] suggested that the thermoelectric figure- of-merit could be greatly enhanced by achieving both high *S* and low $\rho$ in low-dimensional materials which is not possible in the bulk since a low $\rho$ can only be obtained with high carrier densities which in turn leads to a concomitant reduction in *S*.

More specifically, the electronic properties of the $Fe_3O_4$ / $SiO_2$ / p-type Si(100) heterojunction may be exploited for attaining a high thermoelectric figure-of-merit by reducing $\kappa$ and exploiting its high thermoelectric power factor, P¼S2/q. Before elaborating further, we should note that the thermal conductivity of bulk Si is 1.3Wcm$^{-1}$K$^{-1}$ which is about hundred times larger than that of bulk $SiO_2$, i.e., 0.014 Wcm$^{-1}$K$^{-1}$. On the other hand, the thermal conductivity of bulk $Fe_3O_4$ is 0.07 Wcm$^{-1}$K$^{-1}$ and remains more or less constant from the Verwey transition temperature up to 300 K [31]. These values of $\kappa$ change with thickness in the case of thin films. To date there are no investigations on the thermal conductivity properties of $Fe_3O_4$ thin films but the thermal properties of thin $SiO_2$ and Si are still being actively investigated [32] in view of the ongoing downscaling of integrated circuits. It has been found that the thermal conductivity of $SiO_2$ drops down to 0.004 Wcm$^{-1}$K$^{-1}$ for thicknesses of a few nm's [33] while the thermal conductivity of Si is 0.25 Wcm$^{-1}$K$^{-1}$ for a 20 nm thin film of Si [34]. Similarly Si nanowires exhibit a one-hundred fold reduction in thermal conductivity giving very high ZT [3]. Consequently, we anticipate that $Fe_3O_4$ / $SiO_2$ / Si core-shell nanowires or $Fe_3O_4$ on ultra thin $SiO_2$ / Si is very promising indeed for the realization of a high thermoelectric figure-of-merit devices. However, a fundamental understanding of the physical properties and potential of the $Fe_3O_4$ / $SiO_2$ / Si(100) heterojunction is indispensible to the realization of thermoelectric devices with a high performance.

**Summary**

In summary, we have investigated the electrical transport and thermoelectric properties of a $Fe_3O_4$ / $SiO_2$ / *p*-Si(100) heterostructure grown by pulsed laser deposition between 100 and 300K and magnetic fields up to 6 T. We find that the resistance exhibits a sharp drop at high temperatures due to tunneling of electrons from the $Fe_3O_4$ into the *p*-type Si(100). In addition we observe a negative to positive change in the sign of the Seebeck coefficient at the Verwey metalinsulator transition temperature which reaches quite large values (+ 900 $\mu$V/K) close to the $S$ value of the $p$-type Si substrate at 300K as a result of the tunneling of energetic carriers from the $Fe_3O_4$ into the $p$-type Si. The low resistivity of the $Fe_3O_4$ / $SiO_2$ / $p$-Si(100) heterostructure combined with the high value of $S$ in the $p$-type Si, are responsible for the state-ofthe-art values of $P$ (25.5 mW/K$^2$m) obtained at 260K which correspond to underestimated, lower limit values. The large $P$ is attributed to the specific band line up and work function differences between the $Fe_3O_4$ and Si which leads to the formation of an accumulation layer of holes at the Si / $SiO_2$ interface. The use of nm scale Si combined with $Fe_3O_4$ is interesting and promising for the realization of high thermoelectric figure of merit devices.

**Acknowledgements**

The work was supported in part by the University of Cyprus Large Scale Research Project "Thermoelectric Nanosensors Network" by the Cyprus Research Promotion Foundation (project ANAVATHMISI/0609/06), and by the Portuguese Foundation for Science and Technology grant to ICEMS.

**Figure Captions**

**FIG. 1**. Energy band diagram of the $Fe_3O_4$ / $SiO_2$ / *p*-type Si(100) heterostructure in thermodynamic equilibrium including the work functions and electron affinities of $Fe_3O_4$ and *p*-type Si. The Fermi level is at 0 eV. Also shown a schematic diagram of the measurement configuration.

**FIG. 2.** **(a)** *S* vs. *T*, for 200 nm $Fe_3O_4$ / 2 nm $SiO_2$ / *p*-type Si(100) heterostructure. The data for a $Fe_3O_4$ / $AlO_x$ / GaAs structure and a magnetite single crystal [28] are also presented for comparison; **(b)** temperature dependence of the thermoelectric power factor for the same sample. Inset: the temperature dependence of the sheet resistance $R_S$.

**FIG. 3**. Temperature dependence of the Hall voltage, $V_H$, for the 200 nm 200 nm $Fe_3O_4$ / 2 nm $SiO_2$ / *p*-type Si(100) heterostructure. Inset: $V_H(T)$ over an extended temperature range (90–300 K).

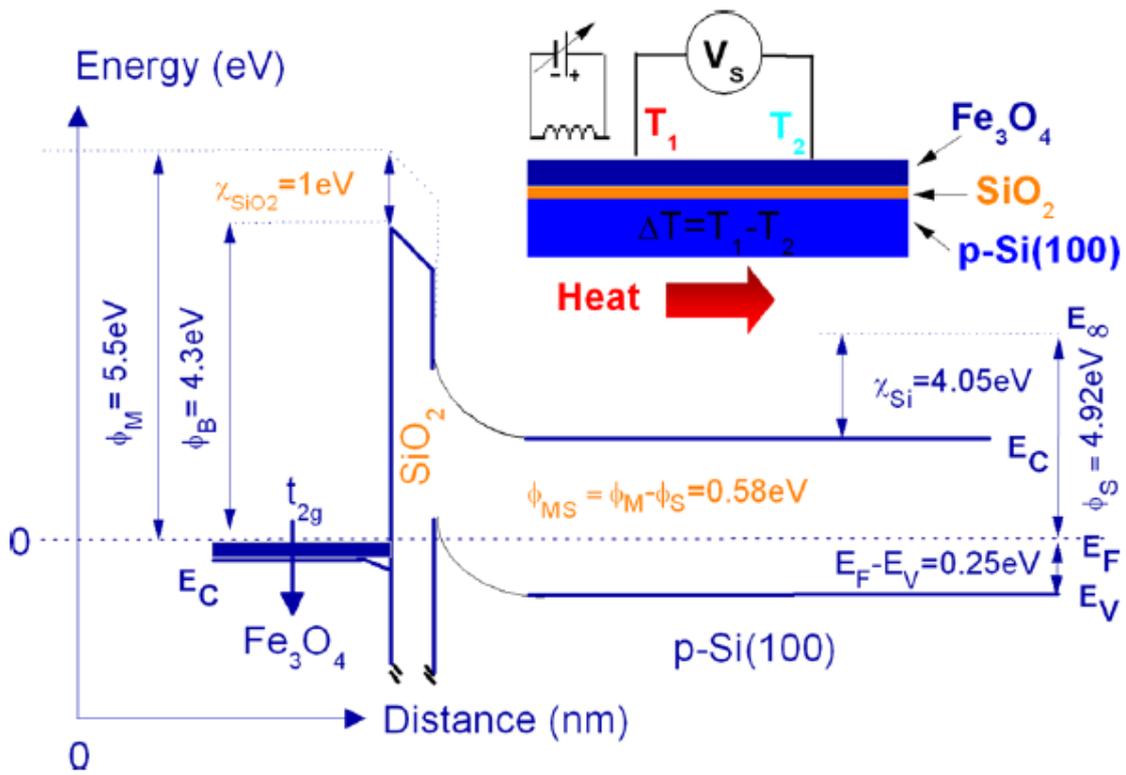

FIG. 1

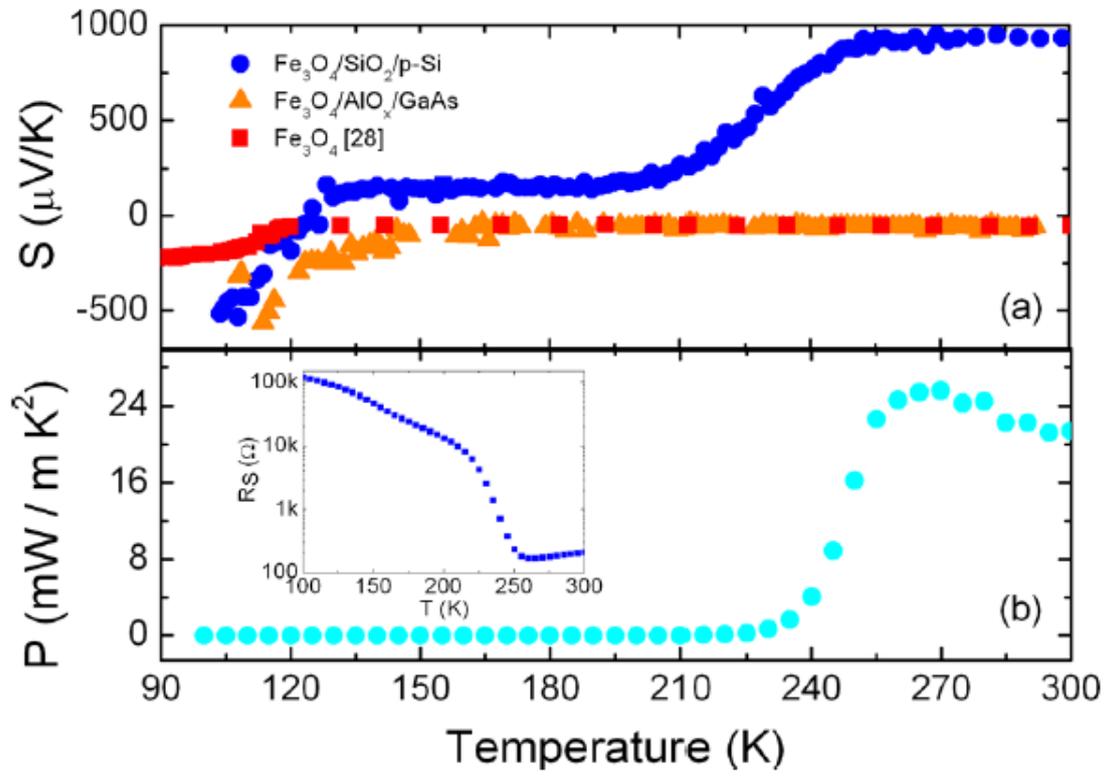

FIG. 2

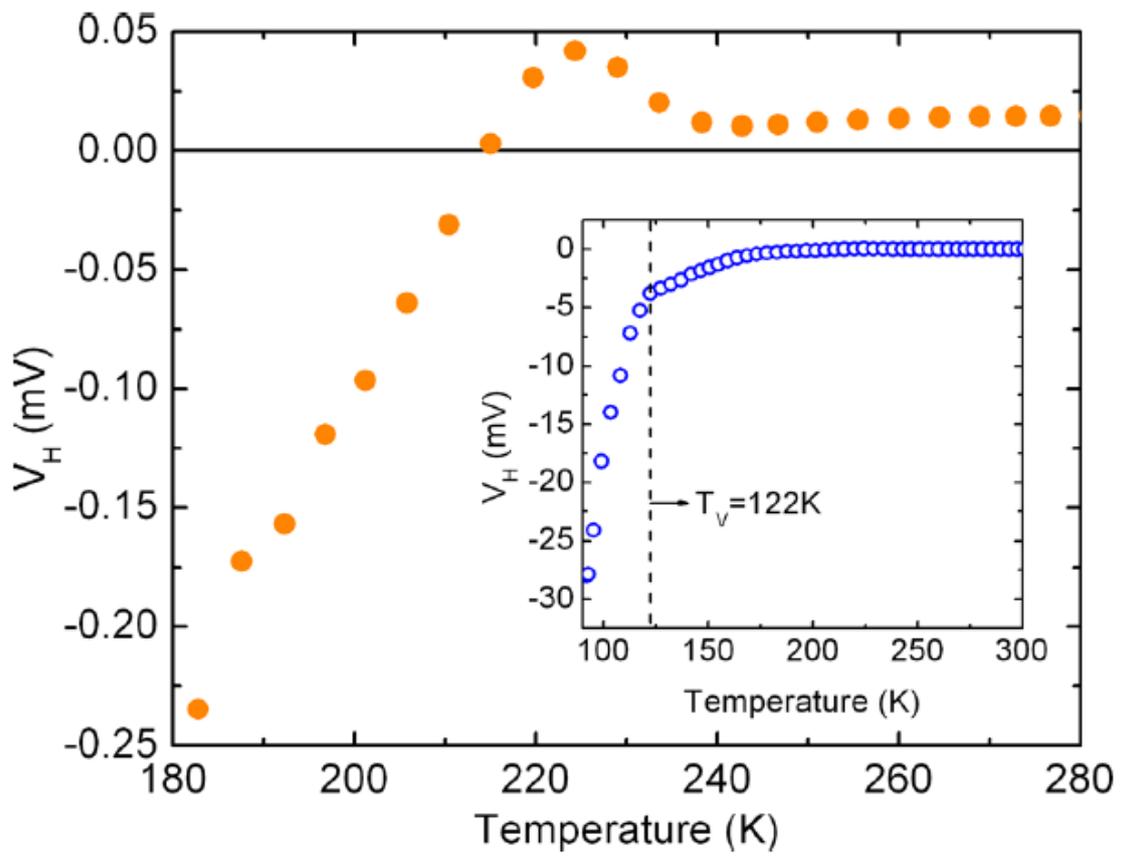

FIG. 3